# A RESOLUTION OF THE VACUUM ENERGY PROBLEM

## Charles B. Leffert

Emeritus Professor, Wayne State University, Detroit, MI 48202
(c_leffert@wayne.edu)

**Abstract**. A new vision of the beginning and expansion of our universe has produced a solution to the vacuum energy problem (also known as "cosmological constant problem"). A new dynamic of cellular spaces and a discrete time has space being produced by a process called spatial condensation (SC). With generic energy defined as Planck's constant times the rate of cellular space production, both the vacuum energy and mass energy contents contribute to the expansion in the ratio $10^{123}/1$, the same ratio of predicted densities by quantum theory and our astronomers. However, unlike mass energy, vacuum energy, like Casimir vacuum energy, does not carry the attribute of mass and so does not gravitate. A geometric derivation of the vacuum energy expansion rate was followed by a second derivation in terms of the evolution of the contents, from radiation to matter to dark mass (not matter). With a new definition of cosmic time, the second derivation was shown to produce exactly the same expansion rate. Free of singularities and inflation, both derivations also produced reasonable values of the cosmological parameters and the second derivation produced a good fit to the supernova Ia data with no acceleration of the expansion rate.

# 1 INTRODUCTION

**1.1 Physics in Trouble**

Recent astronomical measurements [1] have narrowed the average mass-energy density $\rho_m$ of our universe to a value $\Omega_m = \rho_m/\rho_c \approx 0.3$ where $\rho_c = 3H^2/8\pi G$ is the critical density of relativity theory. On the other hand, if the ultraviolet cutoff of vacuum fluctuations is made at the Planck length, quantum mechanics predicts that their energy density is a factor $F \sim 10^{123}$ greater than the astronomer's mass-energy density value of our universe [2]. Because a constant vacuum energy density corresponds to the constant energy density term "lambda" that Einstein added to his general relativity (GR) model of our universe, this embarrassing discrepancy in predicted values has also become known as the "cosmological constant problem." There is a very large literature of theoretical speculation about vacuum energy [3] with little consensus or observational support.

We begin by setting aside all GR global models of our universe. Therefore, the above problem will in the following be called the "vacuum energy problem" because the difference is between a quantum mechanical prediction and astronomical measurements. Theorists point to this well-known problem [4] as indicating something very seriously wrong with present physics.

The view advocated here is that both sciences are correct and so the problem is really a lack of understanding about the nature of *energy*. Most certainly "quantum" *vacuum energy* cannot be the usual gravitating *mass-energy* that astronomers measure. If it is not mass-energy, what other kind of energy could it be? One must begin with:

"What is *energy*?" Then we remember that R. Feynman says in the first volume of his "Lectures on Physics" [5] that: "… we have no knowledge of what energy is." A generic definition of *energy* must be found that includes both vacuum energy and Einstein's mass energy, $E=mC^2$. Other so-called fundamental concepts, such as *space, time* and *mass*, lack a deeper understanding and will also be defined.

The focus for analysis should be directed at finding the correct definition of *energy*; then perhaps with that key discovery, the understanding of other fundamental concepts will fall into place.

Analysis of vacuum energy begins with the proposition that globally the big bang explanation of the observed expansion is wrong; a big bang is not the source of the expansion of our universe. There must be some other, as yet unknown, dynamic at work in our universe that causes it to expand. Knowing the cause of the expansion, we may be able to identify the rate of that dynamic as *energy* at both the macro and the micro level.

As mentioned, quantum mechanics has predicted a very large value for vacuum energy, a factor greater than the mass energy of $F \approx 10^{123}$. A new model of our universe that predicted that same factor F for vacuum energy from other concepts would also produce mutual support for both the quantum prediction as well as for the new model.

## 2 WHY SC-VACUUM ENERGY?--THE VISION

### 2.1 The Missing Dynamic: Spatial Condensation

Edwin Hubble in 1929 was the first to measure the expansion rate of the distances between galaxies in our spatially three-dimensional (3-D) universe. Current theory interprets this dispersion of mass to further conclude that our 3-D space is also increasing and doing so simply by "stretching" (without limits). In contrast, quantum mechanics assigns properties to our 3-D space, including its energy as well as the rapid production, followed by quick annihilation, of virtual particle pairs.

Perhaps the expansion of our 3-D space is due to not just "stretching," but to something that is being produced; the missing dynamic is the "production" of space. It is the author's contention that the source or building blocks for the production of our 3-D space can only be another higher-dimensional space.

A long quest to understand the beginning of our 3-D universe led to the following conception: A preexisting higher-dimensional epi-universe, an m-D epi-space (m>4), underwent a symmetry-breaking event and produced the first very-small hypercube of 4-D space that became a catalytic site for further production of other such 4-D hypercubes.

The edge-length of these tiny 4-D spatial cells were one Planck unit of length, $l_p=1.616 \times 10^{-33}$ cm. Call these cells "Planckton," abbreviated "pk" of number "$N_4$" where each cell produces another, one every Planck unit of time, $t_p=0.539 \times 10^{-43}$ s for a beginning total rate $\dot{N}_4 = dN_4/dt = N_4/t_p$, s$^{-1}$.[*] Also call the much smaller hypercube building blocks of epi-space "m-D pk," and those of our 3-D universe "3-D pk."

Call the process of production of 4-D pk "spatial condensation," where the volume of a 4-D pk in epi-space is less than the total volume of the number of m-D pk that formed it. "Conjecture 1" is that the exposed surface of any foreign object in epi-space will support spatial condensation. The details of spatial condensation have not yet

---
[*] For this paper, consider $dN_4/dt = \Delta N_4/\Delta t$ in the limit $\Delta t = N_t t_p \rightarrow (N_t=1)t_p$

been developed, but in this paper, only the volume of a 4-D pk, $l_p^4$, and their rate of production $\dot{N}_4$, $s^{-1}$ are needed.

To build our three-dimensional (3-D) space and its contents from *space* may seem to be a bold undertaking, but such an attempt is much less bold than the present "new physics" attempt to build our universe from *nothing* [6].

The inrush of the m-D pk to this epi-region of exponential production of 4-D pk drove all of the 4-D pk into a 4-D ball, sometimes called a "4-D core," where its surface became our 3-D universe. Spatial condensation continued, but according to the Conjecture 1, at a much smaller rate (internal 4-D pk no longer exposed), only on the 3-D exposed surface of the 4-D ball and on any mass energy sites produced in its 3-D surface.

Neither *gravity* nor *quantum* behavior were involved in either the "beginning" or "expansion" of our universe.[*] In this new model of our universe, with cellular spaces and discrete times, even *continuum mathematics* is being challenged at the Planck level [7, 8].

The mathematical development of the SC-model for the beginning and expansion of our universe is presented in detail elsewhere [9-14], and a brief summary of the development and its predictions are presented in Appendix A for the convenience of the reader.

From the above one can now define a generic *energy* that can include both vacuum energy and mass energy, because a number rate $\dot{N}$ =dN/dt, $s^{-1}$ times Planck's constant $\hbar$ =1.0546x10$^{-27}$ g cm$^2$ s$^{-1}$ has units of energy,

$$E \equiv \dot{N}\hbar, \text{ g cm}^2 \text{ s}^{-2} \text{ (ergs)}. \tag{1}$$

First, a pictorial representation of spatial condensation during expansion will be presented in § 3 to explain *mass* in the SC-model and the qualitative difference between vacuum energy and mass energy. Then a geometric derivation of vacuum energy is presented in § 4 with comparisons to the mass energy and to Casimir vacuum energy.

The full evolution of the expansion SC-model is summarized in Appendix A, and its prediction of the evolution of vacuum energy in terms of the mass energy contents is discussed in § 5, together with the realization that the geometric and contents derivations are different expressions of one equation. The summary and conclusions are given in § 6.

## 3 WHAT IS SC-VACUUM ENERGY?

### 3.1 Pictorial Representation of Vacuum Spatial Condensation

In Fig. 1, the partial circle represents only one dimension of our 3-D space. However, it could be claimed that the width of the line that forms the circle represents an unknown very small 4-D width of our 3-D universe. In the SC-model, our 3-D universe is the surface of an expanding 4-D ball. The entire 3-D surface of the 4-D ball and its contents of particles of mass are foreign objects in epi-space. According to Conjecture 1, both will support spatial condensation.

The matter particles of our 3-D universe, sketched at the top of Fig. 1, are essentially permanent objects formed along with the 4-D ball and are postulated to have formed persistent columns of arriving m-D pk from epi-space. It is this attribute that we measure as mass where *inertia* is the resistance to change of the angle of the persistent column with respect to **R**. *Dark mass* (at right) consists of a species of 4-D pk that is rejected by the 4-D core but reproduces in place. This dark mass also has permanent existence, and so also supports persistent columns of incoming m-D pk.

---

[*] Quantum mechanics is limited to interactions of matter, radiation and SC-vacuum fluctuations.

On the other hand, the overwhelmingly greater bare surface of the 4-D core consists of reproducing core-acceptable 4-D pk. Each new 4-D pk begins the formation of a persistent column but gets cutoff within one Planck unit of time by its new progeny.[+] This type of spatial condensation is called "bulk spatial condensation" (at left). The overall uniform impact of bulk spatial condensation maintains the general spherical shape of the 4-D ball, i.e., constant 3-D spatial curvature and zero gravity.

Bulk spatial condensation is the overwhelmingly greater contribution to the expansion rate of our universe $\dot{N}_{4\,vac}$, and therefore a much greater vacuum energy $E_{vac} = \dot{N}_{4\,vac} \hbar$ ergs. Einstein's energy, $E=mC^2$, contains the mass symbol (m, g) explicitly, but $E_{vac}$ does not.

Although not indicated in Fig.1, the additional impact of the persistent columns of m-D pk to a large massive object M will produce local curvature (a "dimple") in the 4-D core. In turn, the m-D impact on a nearby test particle m, because of the 3-D curvature caused by M, will drive m towards M, i.e., 3-D gravity (see: Eq. (A.17), Fig. A.1 and [10]).

There is a possible limit on the expansion rate that is suggested from the discussion so far on the SC-dynamics that may be useful in the next § 4.1. The "cellular-discrete time" SC-rate of one new 4-D pk every Planck time suggests a radial velocity of $l_p/t_p=C$. This certainly cannot account for the rate of increase of the radius $\dot{R}$ of a small 4-D ball, but it is a reasonable limit for very large R of a steady-state expansion rate with zero deceleration q=0 ($\ddot{R} \to 0$) or $\dot{R} \to$Constant; call it "Conjecture 2."

## 4 WHERE IS SC-VACUUM ENERGY NOW?

**4.1 SC-Geometric Derivation**

In the SC-geometric derivation, the contents of our 3-D space are ignored to discover what can be derived just from geometric relations and the vision of § 2. The key geometric equations for the volumes of a 4-D ball $V_4$ and its surface $V_3$ are [16],

$$V_4 = \tfrac{1}{2} \pi^2 R^4, \tag{2}$$

$$V_3 = 2\pi^2 R^3. \tag{3}$$

Derivatives with respect to time give the expansion rates (the over-dot represents d/dt),

$$\dot{V}_4 = dV_4/dt = 2\pi^2 R^3 \dot{R} = V_3 \dot{R} = 4V_4 H, \tag{4}$$

$$\dot{V}_3 = dV_3/dt = 6\pi^2 R^2 \dot{R} = 3V_3 H, \text{ where } H=\dot{R}/R, \tag{5}$$

---

[+] Such short-lived columns of m-D pk might still offer some resistance to acceleration of matter.

Volume rates are readily converted to number rates, $N_i = V_i/l_p^i$, $\dot{N}_i = \dot{V}_i/l_p^i$,

$$\dot{N}_4 = \dot{V}_4/l_p^4 = 4N_4 H = \dot{R} N_3/l_p = (N_3/t_p)(\dot{R}/C), \tag{6}$$

$$\dot{N}_3 = \dot{V}_3/l_p^3 = 3N_3 H. \tag{7}$$

The rates of production $\dot{N}_4$ and $\dot{N}_3$ are in units of "planckton per second, pk s$^{-1}$" and are further abbreviated "pks." Useful relations from Eq. (4) are,

$$H = \dot{R}/R = ¼ \dot{V}_4/V_4 = ¼ \dot{N}_4/N_4. \tag{8}$$

Given R as the radius of the 4-D core, $N_{4u}$ and $N_{3u}$ are obtained from Eqs. (2) and (3) respectively, and if one could approximate ($\dot{R}/C$), then Eq. (6) gives $\dot{N}_{4u}$. The vacuum energy of our spatially 3-D universe can then be obtained from Eq. (1),

$$E_{vacu} = \dot{N}_{4u} \hbar = (\hbar N_{3u}/t_p)(\dot{R}/C), \text{ ergs.} \tag{9}$$

Without a further simplifying assumption, the calculation of the vacuum energy of our universe would end here; however it is clear that any details of the evolution of the vacuum energy must be contained in the factor ($\dot{R}/C$).

Before proceeding with a simplifying assumption, there is one more important conclusion that can be extracted from the geometric relations above. If $\dot{V}_3 = 3V_3 H$ from Eq. (5) is universally true, then it must also be true locally in a small local volume $V_3 = (4\pi/3)r^3$. Using Gauss' theorem with a velocity v for the 3-D pk flowing out of an imaginary 2-sphere of radius r, then $\dot{V}_3 = 4\pi r^2 v$ and substituting $3V_3 H$ for $\dot{V}_3$ gives

$$v = Hr. \tag{10}$$

The equivalent of Hubble's law has been derived at the Planck scale for 3-D space itself, and any significant drag on matter content at large distances, or early large H, will have important consequences [10], including influences on the evolution of large-scale structure.

Returning to the simplifying assumption, we try Conjecture 2 for the limit of steady-state expansion, i.e., $\ddot{R}_u = d^2 R_u/dt^2 \to 0$, and further assume that that state is a good approximation for our present universe: If so, then the limiting expansion rate $\dot{R}_u$ is a constant, and Planck's natural units suggest that the constant is the speed of light C = $l_p/t_p$ or ($\dot{R}_{u0}/C$) ≈ 1 or $R_0 ≈ Ct_0$ as in Fig. 2. The subscript zero represents the present.

With this one major assumption, all that is needed to continue is a reasonable estimate of the present radius of the 4-D ball, and that is the same as the radius of our 3-D universe $R_{u0}$. In his 1916 book *Relativity* [17], Einstein had estimated the radius of our closed universe at a value slightly greater than $10^{28}$ cm. My own prior work [10] suggests $R_{u0} = 1.35 \times 10^{28}$ cm and gives,

$$N_{3u0} = 2\pi^2(R_{u0}/l_p)^3 = 1.16 \times 10^{184}, \tag{11}$$

and,

$$H_0 = (\dot{R}_{u0}/R_{u0}) = C/R_{u0} = 2.22 \times 10^{-18} \text{ s}^{-1} = 68.6 \text{ km s}^{-1} \text{ Mpc}^{-1}. \tag{12}$$

Also,

$$t_0 H_0 = t_0(\dot{R}_{u0}/R_{u0}) = t_0 C/Ct_0 = 1, \tag{13}$$

so that,

$$t_0 = 1/H_0 = 1/2.22 \times 10^{-18} \text{ s}^{-1} = 4.50 \times 10^{17} \text{ s} = 14.2 \text{ Gy}. \tag{14}$$

This value of $R_{u0}$ produces the same value for $C/H_0$ that is often considered a natural length scale for our universe. Values of the expansion parameters $t_0$, $H_0$ and $q_0$ are within the range of uncertainty of the astronomer's measurements.

With $(\dot{R}_{u0}/C) = 1$ and $t_p = 0.539 \times 10^{-43}$ s, Eqs. (6) and (9) take the values

$$\dot{N}_{4u0} = N_{3u0}/t_p = 1.16 \times 10^{184}/0.539 \times 10^{-43} = 2.15 \times 10^{227} \text{ s}^{-1}. \tag{15}$$

$$E_{vacu0} = \dot{N}_{4u0} \hbar \approx (\hbar N_{3u0}/t_p) = 2.269 \times 10^{200} \text{ ergs}. \tag{16}$$

Next consider the prediction of Eq. (1) for mass energy using Einstein's equation $E = mC^2$,

$$E_m \equiv \dot{N}_{4m} \hbar = mC^2 = \rho_m N_3 l_p^3 C^2 = (N_3 \hbar/t_p)(\rho_m/\rho_p), \tag{17}$$

where use is made of the Planck units identity $l_p^4 C = \hbar/\rho_p$.[*] Let m be the present mass energy $M_{u0}$ of our universe with average density $\rho_{mu0} \sim 0.3 \rho_c \approx 3 \times 10^{-30}$ g cm$^{-3}$ from recent measurements, and get $E_{m0} \approx 1.3 \times 10^{77}$ ergs or, with $(\dot{R}_{u0}/C) \approx 1$, the present ratio,

$$E_{vac0}/E_{m0} = 1/(\rho_m/\rho_p) = \rho_p/\rho_m \approx 2 \times 10^{123}, \tag{18}$$

in agreement with the quantum mechanics prediction.

Next, let a volume $V_3$ contain both energies, and add for the total energy content

$$E_T = \dot{N}_{4T} \hbar = (\hbar/t_p) N_3 [(\dot{R}/C) + (\rho_m/\rho_p)]. \tag{19}$$

The pre-factor of Eq. (19) has the units of energy $(\hbar/t_p) = 1.956 \times 10^{16}$ g cm$^2$ s$^{-2}$ or ergs.

Dividing total energy by $V_3$ gives the energy density e, ergs/cm$^3$.

$$e_T = E_T/V_3 = \dot{N}_{4T} \hbar/V_3 = (\hbar/l_p^3 t_p)[(\dot{R}/C) + (\rho_m/\rho_p)], \tag{20}$$

---

[*] Radiation energy gravitates, so it should be included under mass-energy in Eq. (17) using $m = E_r/C^2$.

   Warning: This Planck-units identity can convert Eq. (9) to $E_{vac0} = V_{3u0} \rho_p C^2 (\dot{R}/C)$, which is numerically correct but incorrectly implies that vacuum energy is *mass energy*.

where the pre-factor is now a constant and has units of energy density, $(\hbar/l_p^3 t_p) = 4.635 \times 10^{114}$ g cm$^{-1}$s$^{-2}$ or ergs/cm$^3$.[*] Equations (19) and (20) should be correct for any 3-D volume $V_3$ and at any time in the evolution of our universe.

Since a limiting steady-state expansion ($\dot{R}_{u0}/C$) ≈1 is reasonable (Conjecture 2),[+] it is significant that all of these numbers can be derived from such a simple geometric beginning and one major conjecture. This new SC-generic definition of energy is claimed to be the source for the solution to the infamous vacuum energy problem. Indeed, *it is the vacuum energy that drives the expansion of our universe.*

But this solution and our understanding are not yet complete. We have not yet used the contents of our universe to account for the evolution of its vacuum energy density. Before attempting that task, it would be instructive to first check Eq. (19) against the predicted and measured Casimir vacuum energy density.

**4.2 Casimir Vacuum Energy**

Theorists [18] point to the Casimir calculation and experiments as evidence of the existence of a vacuum energy density. Casimir predicted that two parallel plane reflectors, a very small distance apart, would modify the energy of the vacuum field fluctuations in between and cause a small force tending to close the gap. Jaekel, et al. [18] presented the pertinent equations for the force and corresponding energy for the ideal case of two reflectors of area A and distance L apart,

$$F_{cas} = \hbar C\pi^2 A/(240 L^4), \qquad E_{cas} = \hbar C\pi^2 A/(720 L^3), \qquad (J7)$$

and they commented that for A = 1 cm$^2$ and L = 1 μm, $F_{cas}$ ~ 0.1 μN which has been measured. They further noted that even for this ideal case, the force for a given area depends only on the distance L and on two fundamental constants C and $\hbar$.

To compare this calculation and measurement to the prediction of Eq. (19), first multiply numerator and denominator of $E_{cas}$ by L so $V_3 = AL = N_3 l_p^3$, and using $C = l_p/t_p$ and $N_L = L/l_p$, a rearrangement of Eq. (J7) gives for A = 1 cm$^2$ and L = 1 μm,

$$E_{cas} = [\pi^2/(720\, N_L^4)](\hbar N_3/t_p) = f \cdot E_v = 4.33 \times 10^{-7} \text{ erg} \qquad (21)$$

From the volume of the cavity, V = 10$^{-4}$ cm$^3$, $N_3 = 2.37 \times 10^{94}$ and for the present ($\dot{R}/C$ ~1), the normal vacuum energy in the cavity from the first term of Eq. (19), $E_{vac} = 4.635 \times 10^{110}$ ergs. The non-dimensional pre-factor, in square brackets of Eq. (21), is the fraction f = 9.35×10$^{-118}$ of the normal present vacuum energy $E_v$ within the cavity. Note that $F_{cas} = (3/L) E_{cas} = 1.3 \times 10^{-2}$ dyne = 0.13 μN, so Eq. (J7) is in good agreement with the stated measured value of ~0.1 μN for $F_{cas}$. As with "vacuum energy," the concept of "mass" does not appear in the calculation of "Casimir energy"! It is concluded from this geometric calculation that *vacuum energy* does not carry the attribute of mass, E≠mC$^2$. Uniform, *vacuum energy* cannot dimple the 4-D core so *it does not gravitate.*

---

[*] Indeed, $(\hbar/l_p^3 t_p) = \rho_p$, but use of $\rho_p$ would imply *mass energy*.
[+] Unlike current theory, the SC-contents model of Appendix A does not need accelerated expansion.

The space of the "vacuum" cannot "stretch" without limit, as in unlimited "inflation," nor can it carry "potential energy" for conversion to "mass energy" [19].

The next and last task is the development of the SC-contents model to account for the changing vacuum energy during the evolution of our 3-D universe. That task will have to be done in terms of a new *cosmic time* for the changing contents (mass-energy) of our universe and it must predict an expansion rate $\dot{N}_{4u}$ exactly the same as Eq. (6) and $\dot{N}_{4u0}$ approximating Eq. (15).

## 5 WHEN IS VACUUM ENERGY CHANGING?

The spatial condensation process is postulated to be irreversible. Therefore a cosmic time was desired that is asymmetric instead of the symmetric time of present physics. The final desired equation for the evolution of the expansion rate of our universe $\dot{N}_{4u0}$ may finally be expressed as a differential equation with respect to cosmic time. Nevertheless, the usual (local) laws of physics expressed in differential form with respect to the present symmetric, parametric time for R(t) were not considered sufficiently heuristic to lead to the desired equation. Therefore, the attempt was made to define explicitly an asymmetric cosmic time t(R) that itself evolved with the evolution of our universe. The vision continued of cellular spaces and discrete times, but continuum mathematics was certainly deemed justified after the 4-D core was formed.

The early development of this SC-contents model, including its beginning, was presented in two books [9 and 10] and four papers [11-14]. A summary of the SC-expansion model and its predictions is presented in Appendix A.

### 5.1 SC-Contents Derivation

The pertinent equations for the SC-contents derivation are given in Table A.1 of the Appendix. To obtain the value of the expansion rate of our universe from the contents derivation, start with Eq. (A.15) and introduce the speed of light on both sides of the equation to get a useful relation for Eqs. (6), (9), (19) and (20),

$$\dot{R}/C = (R/Ct)(\rho_T/\rho_{T2}). \tag{22}$$

Then use Eq. (8) to convert to 4-D planckton numbers,

$$\dot{N}_{4u} = 4(N_{4u}/t)(\rho_T/\rho_{T2}). \tag{23}$$

This finally is the SC-contents equation that defines the evolution of the expansion rate of our 3-D universe. The goal was that it must predict exactly the same expansion rate of the SC-geometric Eq. (6) and do so over the entire evolution of the expansion. Even though the variables are different, if the goal has been accomplished, one should be able to convert $\dot{N}_{4uC}$ of Eq. (23) to $\dot{N}_{4uG}$ of Eq. (6).

First, note from Eq. (A.15) that Eq. (23) can be written as $\dot{N}_{4uC}=4N_{4uC}(\dot{R}/R)$, and further using Eq. (8) it can be written as $\dot{N}_{4uC}=(RN_{3u}/Ct_p)(\dot{R}/R)$, which reduces to

$$\dot{N}_{4\,uC} = (N_{3u}/t_p)(\dot{R}/C) = \dot{N}_{4\,uG}, \quad \text{Q.E.D.} \tag{24}$$

From Eq. (22), when in the limit $(\rho_T/\rho_{T2})\to 1$, so also does $(\dot{R}/C)\to 1$. From scaling Eqs. (A.9) to (A.11), it is clear that the new scaling ($\propto R^{-2}$) of dark mass (with the new definition of cosmic time) made the dual limit possible.

Equation (23) times $\hbar$ also gives the same vacuum energy $E_{vac}$ as Eq. (16) and divided by $V_3$ the same vacuum energy density $e_{vac}$, and in particular, the evolution of both with the expansion of our universe (see Fig. 3 and Fig. 4). Generic energy might have been defined as Planck's constant times a different rate $\dot{L}$, s$^{-1}$, but as Fig. 4 shows, only $E_{vac} = \dot{N}_{4u}\hbar$ provides correspondence with the beginning of spatial condensation. For the photon, $\dot{L}$ = frequency, $\nu$. $E_{Ph} = 2\pi\hbar\nu = \dot{N}_{4\,Ph}\hbar$, ergs, so $\dot{N}_{4\,Ph} = 2\pi\nu = \omega$.

**5.2 Equality of Limits as t → ∞**

Steady state expansion with $\dot{R}/C = \rho_T/\rho_{T2} = 1$ and Eq. (23) = Eq. (5) gives

$$N_{4u} = \tfrac{1}{4} N_{3u}(t/t_p) = \tfrac{1}{4} N_{3u}N_t, \tag{25}$$

where $N_t$ represents the total ticks at time t of a hypothetical Planck clock that made one tick every Planck second (~$10^{-43}$ s) since the very first 4-D pk was produced. Integer Eq. (25) does not involve rates but integer numbers of pk and Planck time according to the claim of cellular spaces and discrete times. Surprisingly, if Eq. (25) is returned to normal units (R,t), it translates to R=Ct that was postulated for the limit t→∞ in the development of the SC-model. Equation (25) says that at any time t, the number of 4-D pk in the 4-D core is equal to one quarter the number of 3-D pk in our 3-D universe times the past number of ticks of the Planck clock.

# 6 CONCLUSIONS

Instead of our 3-D space *stretching* in order to expand, a new vision of our beginning involves a new dynamic of spatial condensation that actually *produces* our expanding space. This new vision has our spatially 3-D closed universe as the surface of an expanding 4-D ball embedded in a pre-existing m-D epi-space. A geometric derivation of the expansion rate of our universe, ignoring its mass energy contents, was shown equal to the expansion rate of a derivation in terms of the evolution of its contents.

This new cellular-space, discrete-time cosmological model predicts reasonable values for the cosmological parameters and provides a solution of the well-known vacuum energy problem. Planck's constant times the cellular expansion rate of the 4-D ball gives a vacuum energy that is a factor $10^{123}$ times the contribution due to the mass energy contents of our universe in agreement with the prediction of quantum theory. Also, in agreement with Casimir vacuum energy, the SC-model predicts vacuum energy does not carry the attribute of *mass*, $E_{vac} \neq mC^2$, and does not gravitate, in conflict with present inflation theory.

Once produced, the 4-D pk of the 4-D core are certainly conserved, and that certainly negates any quantum annihilation and creation of universes [20].

# 7 ACKNOWLEDGEMENTS

The author thanks his good friend, Emeritus Professor Robert A. Piccirelli, for extensive discussions of the new physical concepts.

# APPENDIX A

## Summary of the SC-Contents Model [10]

**A.1 The Cosmological Model**

The scale factor R has units of length for our 3-sphere, spatially 3-dimensional expanding universe; G is the gravitational constant; C is the speed of light; and H is the Hubble parameter. Present values have subscript 0 and cgs units are assumed. Other subscripts include: T=total, r=radiation, m=matter and x=dark mass (not dark matter). Pertinent equations of the new model [hereafter: "SC-model"] are listed in Table 1.

Table A.1  Derivation of Model

| | | |
|---|---|---|
| Universal constant: | $\kappa = Gt^2\rho_T = Gt_0^2\rho_{T0} = 3/32\pi$. | (A.1) |
| From $T_0$=2.726 K: | $\rho_{r0} = 9.40 \times 10^{-34}$. | (A.2) |
| From nucleosynthesis: | $\rho_{m0} = 2.72 \times 10^{-31}$. | (A.3) |
| Limiting expansion rate: | $(dR/dt)/C \to 1$ as $t_0 \to \infty$. | (A.4) |
| Present age, (Input): | $t_0$ | (A.5) |
| From (A.1): | $\rho_{T0} = (\kappa/G)/t_0$. | (A.6) |
| From (A.6): | $\rho_{x0} = \rho_{T0} - \rho_{r0} - \rho_{m0}$. | (A.7) |
| Redshift Z (Input): | $(1+Z) \equiv R_0/R$ | (A.8) |
| Radiation at Z: | $\rho_r = \rho_{r0}(1+Z)^4$. | (A.9) |
| Matter at Z: | $\rho_m = \rho_{m0}(1+Z)^3$. | (A.10) |
| Dark Mass at Z: | $\rho_x = \rho_{x0}(1+Z)^2$. | (A.11) |
| Total at Z: | $\rho_T = \rho_r + \rho_m + \rho_x$. | (A.12) |
| Cosmic time: | $t(R) = + (t_0^2 \rho_{T0}/\rho_T(R))^{1/2}$. | (A.13) |
| From derivatives, d/dt: | $\rho_{T2} = 2\rho_r + 3/2\, \rho_m + \rho_x$. | (A.14) |
| From derivatives, d/dt: | $H = \dot{R}/R = (\rho_T/\rho_{T2})/t$, | (A.15) |
| and | $H_0 = \dot{R}_0/R_0 = (\rho_{T0}/\rho_{T20})/t_0$. | (A.16) |

The scaling with the expansion of radiation, Eq. (A.9), and matter, Eq. (A.10), are borrowed from the big bang model, as is the value of $\kappa$ for early Friedmann radiation.

The postulated scaling, Eq. (A.11), of the new and now dominant stuff called "dark mass," is the key signature of this new cosmological model. Its density decreases with the expansion but its total mass, always in individual clumps, increases with the expansion. It is not a 3-D substance and so does not interact with radiation or matter except gravitationally, where it certainly contributes to the curvature of 3-D space. The distribution of these miniscule dark mass seeds at the beginning of the expansion sets the pattern for the present large-scale structure, including voids, and contributes to the early formation of black holes and fit to supernova Ia data for $t_0$=13.5 Gy (see Figs. 5 and 6) and no acceleration of the expansion rate.

The basic postulate for cosmic time, Eq. (A.13), was made in terms of partial times $\Gamma_i$ where $t^{-2} = \sum_i \Gamma_i^{-2}$ and $\Gamma_i = (\kappa/G)/\rho_i(Z)$ where $\rho_i$ are given by Eqs. (A.9) to (A.11). Cosmic time begins with value $t = \infty$ and jumps to $t=t_p$ with the first 4-D pk. With age set to $t_0$=13.5 Gy, the SC-model predicted the following values for the present cosmological parameters: $R_0$=1.354x10$^{28}$ cm, $H_0$=68.6 km s$^{-1}$ Mpc$^{-1}$, $\Omega_B$=0.031, $\Omega_{DM}$=0.248, $\Omega_{DM}/\Omega_B$=8.0, ($\dot{R}/C$)=1.005 and $q_0$=0.0084 (i.e., approaching steady-state expansion), all within the range of uncertainty of our astronomer's measurements.

**A.2 SC-Gravity only Appears Attractive [10]**

Shown to support the m-D pk source of *mass* that is missing in vacuum energy.

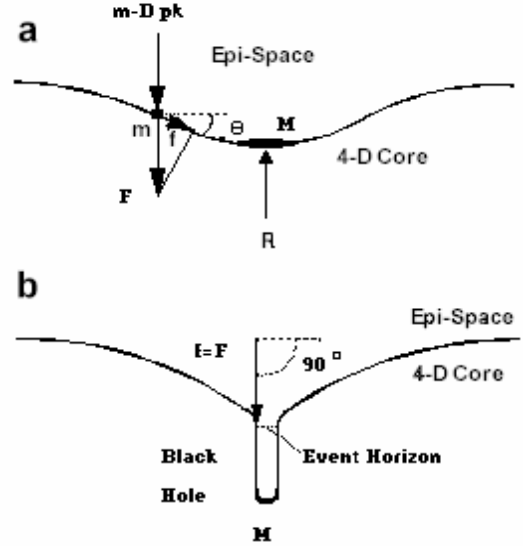

Fig. A.1 (a) Columns of arriving m-D particles to large mass M (not shown) dimple the 4-D core. Columns of arriving m-D particles to test particle m at 3-D radius r produce a 4-D radial force F on m with r-component $F \sin\theta = GmM/r^2$ toward M.

First, divide by m to work with accelerations: $a_r = F_m/m \sin\theta = GM/r^2$. Can $a_R = F_m/m$ and $\sin\theta$ be separated to calculate $F_m/m$? (b) Yes! $\sin\theta = 1$ at the event horizon $R_s$ of a black hole of mass M to give the first parameter $a_R = F_m/m$ ever calculated of the epi-universe.

(c) Vacuum energy does not dimple the 4-D core to "attract" other mass: nor does it have permanence (mass) to respond to the curvature produced by a massive object.

Persistent columns of m-D pk from epi-space are the source of *mass* and accelerations $a_R$ and $a_r$.

Let $N_M = M/m_p$, $N_r = r/l_p$, and for $R_s = 2GM/C^2$, $N_r = 2N_M$ so,

$a_r = F_m/m \sin\theta = -\xi(N_M/N_r^2)$ where $\xi \equiv (C^2/l_p) = 5.569 \times 10^{53}$ cm s$^{-2}$. (A.17)

At $r = R_s$, $a_R = F_m/m = -GM/R_s^2 = -\xi/4 (1/N_M)$. (A.18)

Eqs. (A.17)/(A.18) give $\sin\theta = 4(N_M/N_r)^2 = \chi(M^2/r^2)$, (A.19)

where: $\chi \equiv 4(l_p/m_p)^2 = 2.204 \times 10^{-56}$ cm$^2$ g$^{-2}$. (A.20)

On m, Relative to M:
Sun: $a_R = 1.51 \times 10^{15}$ cm s$^{-2}$, $\theta \approx \sin\theta = 3.94 \times 10^{-16}$ rad., $a_r = 0.597$ cm s$^{-2}$
Earth: $a_R = 5.08 \times 10^{20}$ cm s$^{-2}$, $\theta \approx \sin\theta = 1.93 \times 10^{-18}$ rad., $a_r = 980.$ cm s$^{-2}$

Negative "gravitational energy" does not exist in 3-D space; it is a "potential" energy instantly available from epi-space to a mass m free to respond to a 3-D curvature of $\sin\theta$.

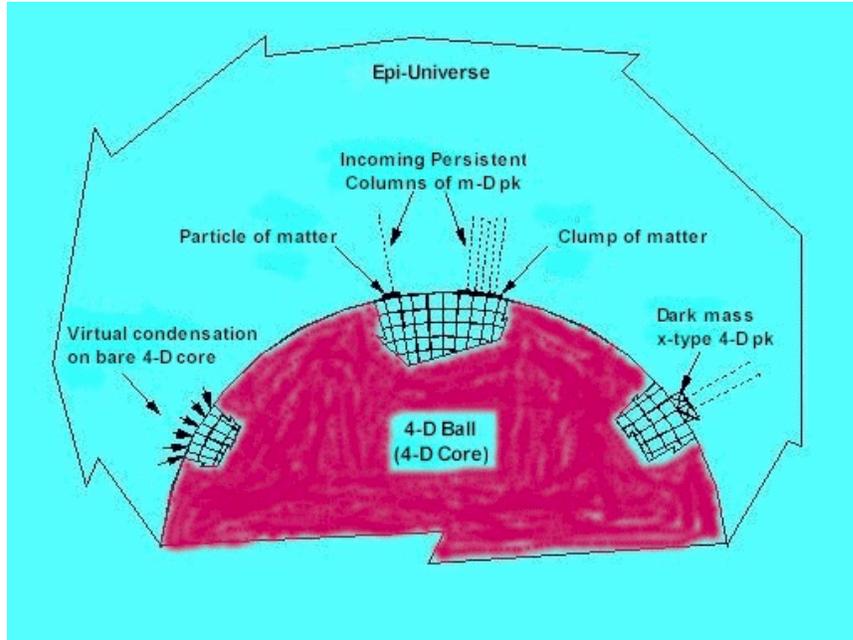

Fig. 1 Spatial condensation to the contents of the surface of the 4-D core is shown for matter and dark mass (mass-energy) by incoming persistent columns of m-D pk. Bulk spatial condensation (vacuum energy) to the bare 4-D core vacuum fluctuations can be probed by Casimir-type experiments.

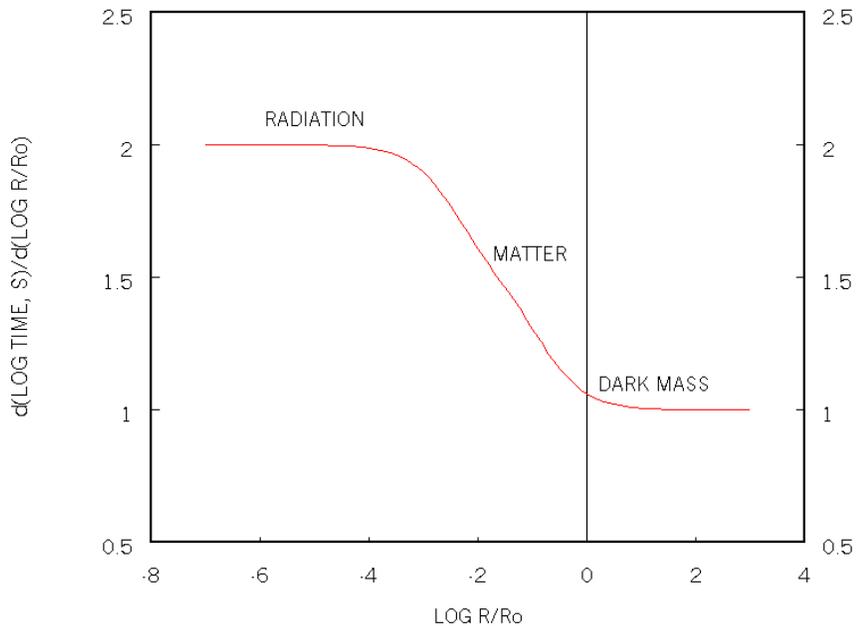

Fig. 2 Time is unique in the SC-model in that it is defined in terms of the changing resistance to spatial condensation as the dominant mass energy evolves with the expansion from radiation to matter and finally to dark mass. The key equation can also be written as $dt/t = k\, dR/R$ or $d\log t/d\log R = k$ where $k = \rho_{T2}/\rho_{T1}$ varies from 2 to 3/2 to 1 with the expansion.

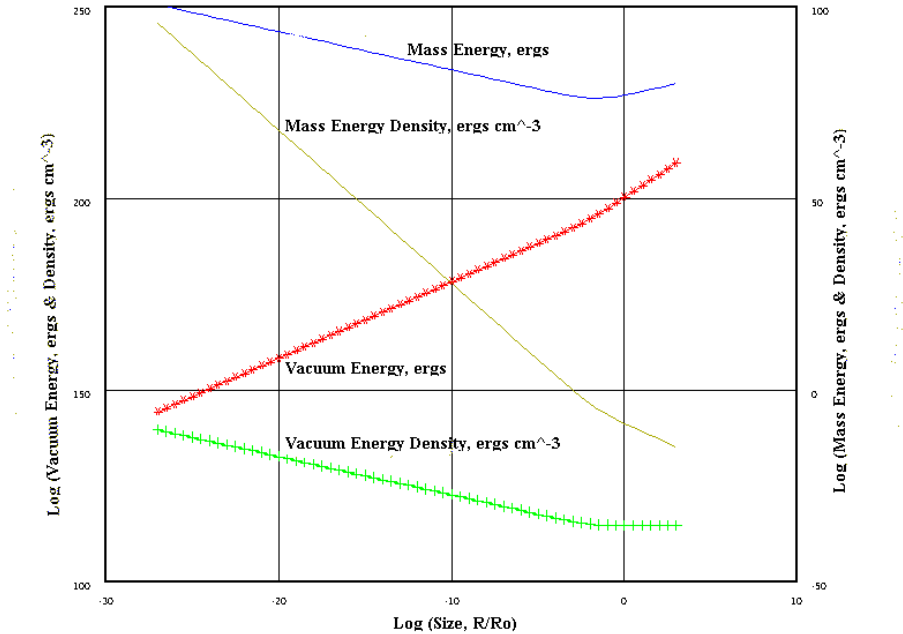

Fig. 3 Evolution of both vacuum and mass energy, and their densities, are shown over the entire past of our 3-D universe since compaction of the 4-D core. At present, $R/R_0=1$, the vacuum energy density has almost reached its constant value of $4.635 \times 10^{114}$ ergs cm$^{-3}$. Only the vacuum energy $\dot{N}_4 \hbar$ extends back through the "free 4-D pk" *beginning* (see Fig. 4).

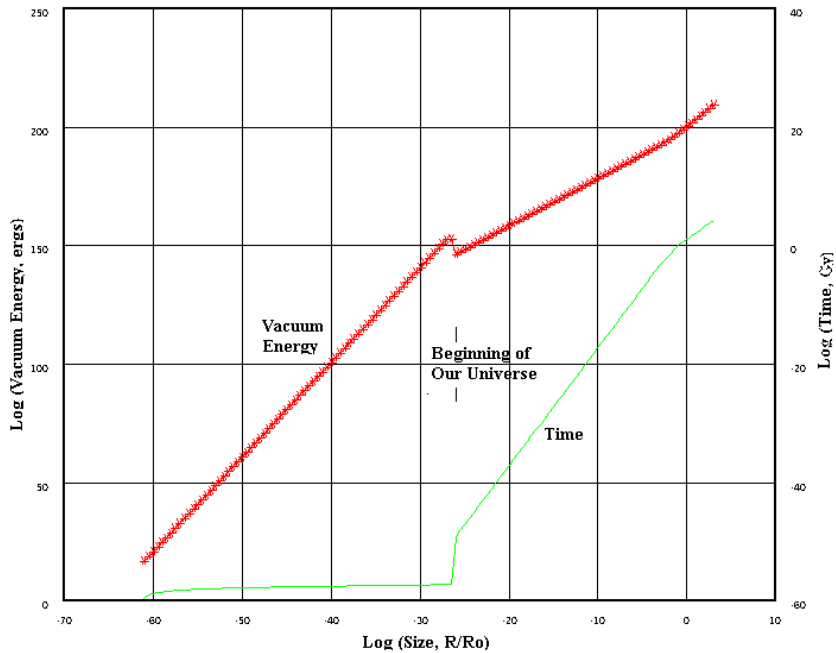

Fig. 4 The vacuum energy curve $E_v(R)$ and cosmic time $t(R)$, are extended back through the *beginning* to the symmetry-breaking event in epi-space of the production of the first 4-D pk of dark mass. During the *beginning* production of "free" 4-D pk, until compaction of the 4-D core, $R = (2N_4/\pi^2)^{1/4} l_p$.

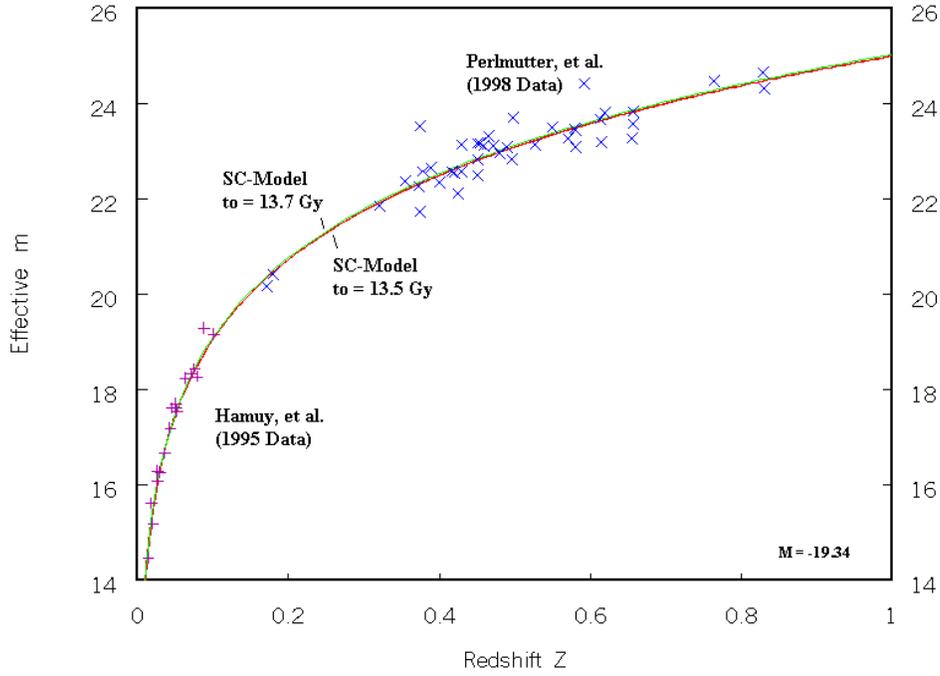

Figure 5 [SNIa Data: 21]

| | | |
|---|---|---|
| $\Omega_{B0}$ = 0.031 | $t_0$ = 13.5 Gy | $q_0$ = 0.0084 |
| $\Omega_{DM0}$ = 0.248 | $H_0$ = 68.6 km s$^{-1}$ Mpc$^{-1}$ | $(\dot{R}/C)_0$ = 0.005 |
| $\Omega_{mass0}$ = 0.279 | $(\Omega_{DM}/\Omega_B)_0$ = 8.05 | $R_0$ = 4388 Mpc |

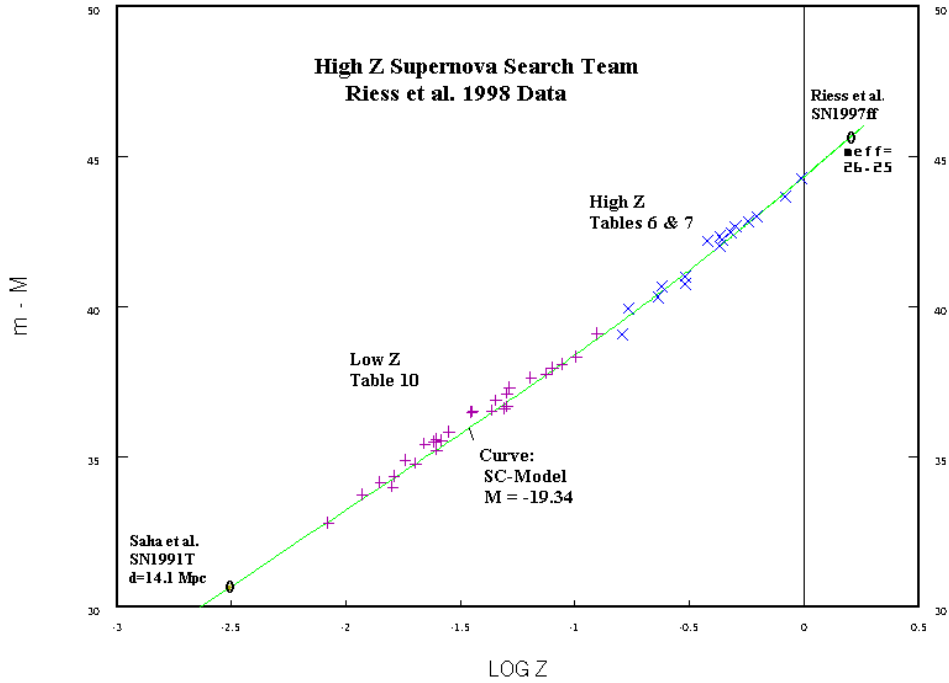

Figure 6 [SNIa Data: 22]